\documentclass[aps,prl,onecolumn,epsfig,amsmath,amstex,floatfix]{revtex4}

\usepackage{epsfig,amsmath}
\usepackage{graphicx}
\usepackage{dcolumn}
\usepackage{bm}
\usepackage{bbold}
\newcommand{\beq}{\begin{equation}}
\newcommand{\eeq}{\end{equation}}
\newcommand{\beqa}{\begin{eqnarray}}
\newcommand{\eeqa}{\end{eqnarray}}

\newcommand{\la}{\langle} 
\newcommand{\ra}{\rangle}

\def\pla#1{{ Phys.\ Lett. A\/} {\bf#1}}
\def\pra#1{{ Phys.\ Rev. A\/} {\bf#1}}

\def\prl#1{{ Phys.\ Rev.\ Lett.} {\bf#1}}

\begin{document}
\title {Wave-Particle Duality Controlled by Single-Photon Self-Entanglement}
\author{X.-F. Qian$^{1}$}
\email{xiaofeng.qian@stevens.edu}
\author{K. Konthasinghe$^{2,3}$}
\author{S. K. Manikandan$^{2,4}$}
\author{D. Spiecker$^{5}$}
\author{A. N. Vamivakas$^{2,3,4}$}
\author{J. H. Eberly$^{2,3,4}$}

\affiliation{
$^{1}$Department of Physics, Stevens Institute of Technology, Hoboken, NJ 07030, USA\\
$^{2}$Center for Coherence and Quantum Optics,
University of Rochester,
Rochester, New York 14627, USA\\
$^{3}$The Institute of Optics, University of Rochester, Rochester, NY 14627, USA\\
$^{4}$Department of Physics \& Astronomy, University of Rochester,
Rochester, New York 14627, USA\\
$^{5}$National Technical Institute for the Deaf, Science and Mathematics Department, Rochester Institute of Technology,
Rochester, New York 14623, USA}

\date{\today }

\begin{abstract}
We experimentally observe that quantum duality of a single photon is controlled by its self-entanglement through a three-way quantum coherence identity $V^2+D^2+C^2=1$. Here V, D, C represent waveness, particleness, and self-entanglement respectively. 
\end{abstract}


\maketitle

Wave-particle duality has been regarded in many discussions as the {\em only} mystery of quantum mechanics \cite{Feynman}. According to de Broglie \cite{deB} and Bohr \cite{Bohr-28}, a quantum particle or quantic entity contains both wave and particle characters, but the observation of one character automatically denies the other one. To comprehend such confusing facts, Bohr proposed the principle of ``Complementarity" \cite{Bohr-28}, which states (as reinforced by Bohr in 1949 \cite{Schilpp}), ``...  evidence obtained under different conditions {\em cannot be comprehended within a single picture}, but must be regarded as complementary in the sense that only {\em the totality of the phenomena exhausts the possible information} about the objects". Two important messages must be emphasized from Bohr's statement. First, one can conclude from the italic phrases, as Whitaker paraphrases Murdoch \cite{Whitaker}, that two factors are critical and must be involved in any correct complementary description - {\em mutual exclusivity} and {\em joint completion}. Second, Bohr's description doesn't restrict complementarity to only two conflicting characters.  

Quantitative investigations of quantum duality were not initiated until about 50 years later by Wootters and Zurek \cite{Wootters-Zurek-79} in 1979 and then followed by many others \cite{others}, leading to a simple inequality between interference visibility $V$ (wave property) and distinguishability $D$ (particle property), i.e., $V^2+D^2 \le 1$. However, such an inequality doesn't guarantee the decreasing of one quantity with the increasing of the other. Thus, it cannot represent {\em mutual exclusiveness}. Moreover, the inequality allows the extreme case $V=D=0$ to exist, indicating that the quantum entity is neither a wave nor a particle, but doesn't reveal its residual information. Something must be missing. So the inequality cannot embody the {\em joint completion} requirement either. 

As a resolution of these two issues, here we briefly report our recent single-photon self-interference experiment \cite{VDC} examining quantitatively the messages from Bohr's complementarity statement. We observe how single photon self-entanglement serves the purpose of completing the duality inequality with a three-way coherence identity \cite{QVE}, 
\beq \label{VDC1}
V^2+D^2+ C^2 =1, 
\eeq
where the self-entanglement measure $C$ is concurrence \cite{Wootters}. We point out that a similar identity relation was obtained in the context of two qubits \cite{Bergou}. But here we take the conventional view that quantum complementarity is about the intrinsic nature of a single quantum entity itself.


\begin{figure}[h!]
\includegraphics[width=9cm]{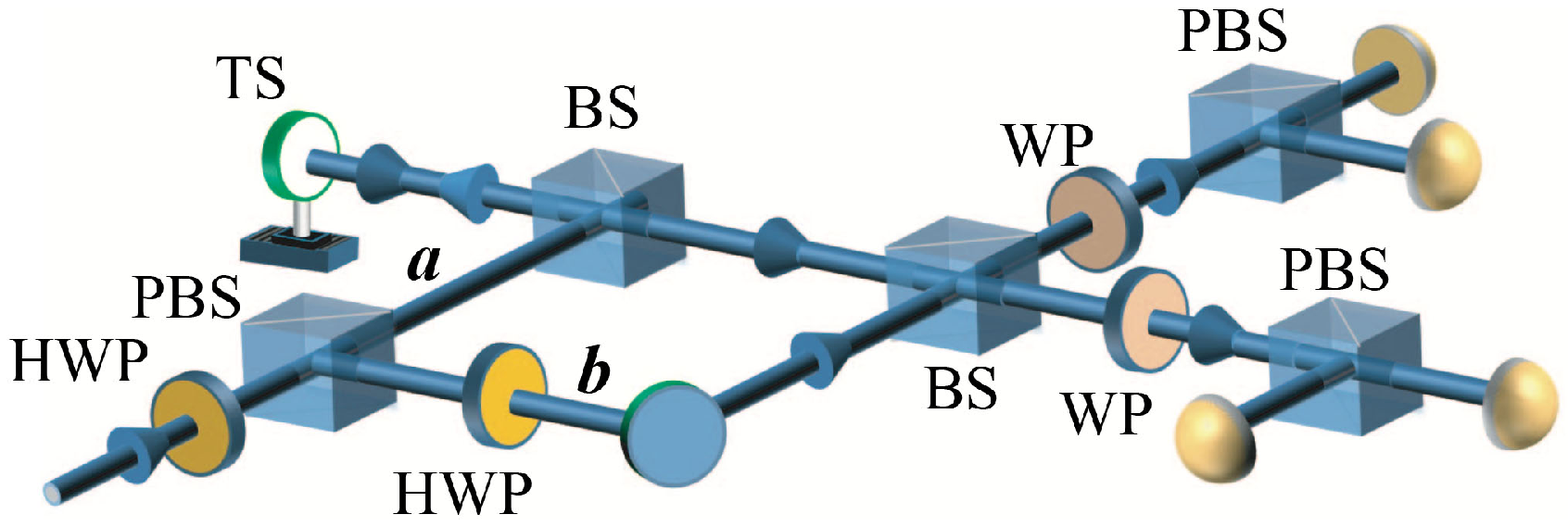}
\includegraphics[width=4cm]{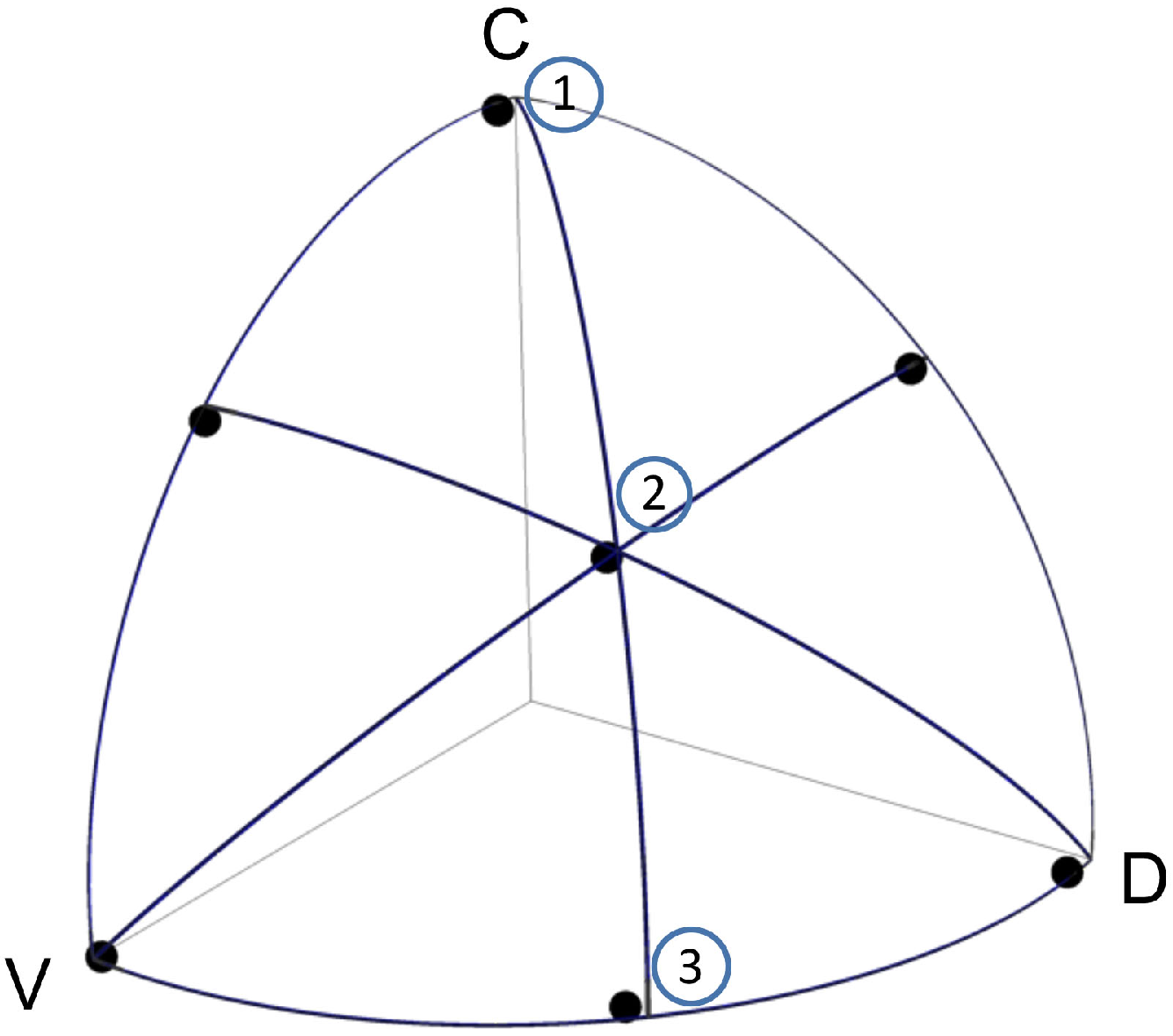}
\caption{Left: schematic setup. Right: measured $V,D,C$ values (black dots) on the unit sphere. } 
\label{sphere}
\end{figure}

An analog of the double-slit scenario is illustrated by the Mach-Zehnder interferometer in the left panel of Fig.~\ref{sphere}, where identical single photons approach the two paths $a$, $b$ and combine at the second beamsplitter (BS). The two-mode single photon field can be written as $\hat{E}^{(+)}=\hat{E}_a^{(+)}+\hat{E}_b^{(+)}=a_a e^{i\varphi_a}+a_be^{i\varphi_b} $ which characterizes the positive frequency part. Here $a_a$ and $a_b$ are the photon annihilation operators for modes $a$ and $b$ respectively, $\varphi_a$ and $\varphi_b$ indicate the phases respectively associated with the propagation in paths $a$ and $b$. The single photon state can be generally described as 
\beq
|\Psi\ra = c_a|\mathbb{1}_a\ra| \otimes \phi_a\ra + c_b|\mathbb{1}_b\ra \otimes |\phi_b\ra, \label{state}
\eeq
where $c_a$, $c_b$ are normalized coefficients with $|c_a|^2+|c_b|^2=1$, and $|\mathbb{1}_a\ra$, $ |\mathbb{1}_b\ra$ are single photon mode states indicating respectively one photon in modes $a$, $b$ and no photon elsewhere. Here $|\phi_a\ra$ and $|\phi_b\ra$ are two corresponding normalized states of all the remaining intrinsic degrees of freedom of the single photon, with generic partial correlation $\gamma = \la \phi_a|\phi_b\ra$ and  $0\le |\gamma| \le 1$. Then by convention the wave property of the photon is quantified by self-interference visibility $V$ and the particle property is quantified by which-way distinguishability $D=|p_a-p_b|$ where $p_{a}$,  $p_{b}$ are the probabilities of the photon taking paths $a, b$. The self-entanglement or non-separability between path (described by modes $|\mathbb{1}_{a}\ra$, $|\mathbb{1}_{b}\ra$) and the remaining all internal degrees of freedom (ascribed to the state functions $|\phi_a\ra$, $|\phi_b\ra$) demonstrated in state (\ref{state}) can be quantified with concurrence $C$ after Schmidt decomposition. All three quantities can be obtained as 
 \beqa
V=2|c_ac_b\gamma|, \quad\quad D=\sqrt{1-4|c_ac_b|^2}, \quad\quad   C=2|c_ac_b|\sqrt{1-|\gamma|^2}.
\eeqa
It is straightforward to show that these quantities satisfy the three-way coherence identity given in (\ref{VDC1}).

We have tested the analytical result experimentally with single photons generated by a hexagonal boron nitride quantum dot. Single photons pass through a modified Mach-Zehnder interferometer and a tomographic setup as shown in the left panel of Fig. \ref{sphere}. Intrinsic degrees of freedom of the photon in terms $|\phi_a\ra$, $|\phi_b\ra$ are represented by two different polarization states $|s_a\ra$, $|s_b\ra$ in the experiment. Seven different measured $(V,D,C)$ values, corresponding to seven different single photon states, are shown by the black dots in Fig.~\ref{sphere}. If the single photon state is tuned so that its $(V,D,C)$ value evolve from points 1 (via 2) to 3 as shown on the sphere, then both wave property (visibility $V$) and particle property (distinguishability $D$) are decreasing. This confirms the violation of the {\em mutual exclusiveness} criteria of the complementarity principle by the $VD$ inequality $V^2+D^2\le1$. Also, point 1 shows the measurement of the extreme case of $V=D=0$, confirming the violation of the {\em joint completion} requirement by the $VD$ inequality. Apparently, when self-entanglement $C$ is considered, both {\em mutual exclusiveness} and {\em joint completion} are satisfied. In fact, all the black dots confirm the identity (\ref{VDC1}) connecting the intrinsic single-photon properties, i.e., visibility, distinguishability and self-entanglement, indicating the complementary behavior of three rather than two characters. 

We acknowledge financial support from DARPA D19AP00042, and NSF grants PHY-1203931, PHY-1505189, and INSPIRE PHY-1539859.

\end{document}